\documentclass[letter,longauth]{aa}
\usepackage{graphicx}
\usepackage[]{natbib}
\usepackage{multirow}
\usepackage{amsmath} 
\usepackage{txfonts}  
\usepackage{enumitem}
\usepackage[colorlinks=true,linkcolor=blue,citecolor=blue]{hyperref}  

\def\deg{$^{\circ}$}
\def\degb{^{\circ}}

\newcommand{\pc}{\,\hbox{pc}}

\begin{document}

\title{Investigating point sources in MWC 758 with SPHERE \thanks{Based on data collected at the European Southern Observatory, Chile under programs 096.C-0241 and 1100.C-0481}
}

\author{        
A. Boccaletti\inst{\ref{lesia}}
\and E. Pantin\inst{\ref{cea}}
\and F. M{\'e}nard\inst{\ref{ipag}}
\and R. Galicher\inst{\ref{lesia}}
\and M. Langlois\inst{\ref{lam},\ref{cral}} 
\and M. Benisty\inst{\ref{ipag}}
\and R. Gratton\inst{\ref{inaf}}
\and G. Chauvin\inst{\ref{ipag}}
\and C. Ginski\inst{\ref{amsterdam},\ref{leiden}}
\and A.-M. Lagrange\inst{\ref{lesia},\ref{ipag}}
\and A. Zurlo\inst{\ref{santiago},\ref{escuela}}
\and B. Biller\inst{\ref{edin}}
\and M. Bonavita\inst{\ref{inaf},\ref{edin}}    
\and M. Bonnefoy\inst{\ref{ipag}} 
\and S. Brown-Sevilla\inst{\ref{mpia}} 
\and F. Cantalloube\inst{\ref{lam}}
\and S. Desidera\inst{\ref{inaf}} 
\and V. D'Orazi\inst{\ref{inaf}} 
\and M. Feldt\inst{\ref{mpia}} 
\and J. Hagelberg\inst{\ref{ipag}} 
\and C. Lazzoni\inst{\ref{inaf}}  
\and D. Mesa\inst{\ref{inaf}} 
\and M. Meyer\inst{\ref{eth},\ref{michigan}}
\and C. Perrot\inst{\ref{lesia},\ref{valpa},\ref{nucleo}} 
\and A. Vigan\inst{\ref{lam}}
\and J.-F. Sauvage\inst{\ref{onera}}
\and J. Ramos\inst{\ref{mpia}}
\and G. Rousset\inst{\ref{lesia}}
\and Y. Magnard\inst{\ref{ipag}} 
}
 
\institute{
LESIA, Observatoire de Paris, Universit{\'e} PSL, CNRS, Sorbonne Universit{\'e}, Univ. Paris Diderot, Sorbonne Paris Cit{\'e}, 5 place Jules Janssen, 92195 Meudon, France\label{lesia}
\and CEA, IRFU, DAp, AIM, Universit\'e Paris-Saclay, Universit\'e Paris Diderot, Sorbonne Paris Cit\'e, CNRS, F-91191 Gif-sur-Yvette, France\label{cea}
\and Univ. Grenoble Alpes, CNRS, IPAG, F-38000 Grenoble, France\label{ipag}
\and Aix Marseille Universit{\'e}, CNRS, LAM (Laboratoire d’Astrophysique de Marseille) UMR 7326, 13388 Marseille, France\label{lam}
\and CRAL, UMR 5574, CNRS, Universit{\'e} de Lyon, Ecole Normale Sup{\'e}rieure de Lyon, 46 All{\'e}e d’Italie, F–69364 Lyon Cedex 07, France\label{cral}
\and INAF-Osservatorio Astronomico di Padova, Vicolo dell’Osservatorio 5, I-35122 Padova, Italy\label{inaf}
\and  Anton Pannekoek Institute for Astronomy, University of Amsterdam, Science Park 904,1098XH Amsterdam, The Netherlands\label{amsterdam}
\and  Leiden Observatory, Leiden University, 2300 RA Leiden, The Netherlands\label{leiden}
 \and N\'ucleo de Astronom\'ia, Facultad de Ingenier\'ia y Ciencias, Universidad Diego Portales, Av. Ejercito 441, Santiago, Chile\label{santiago}\\
 \and Escuela de Ingenier\'ia Industrial, Facultad de Ingenier\'ia y Ciencias, Universidad Diego Portales, Av. Ejercito 441, Santiago, Chile\label{escuela}\\
\and SUPA, Institute for Astronomy, The University of Edinburgh, Royal Observatory, Blackford Hill, Edinburgh, EH9 3HJ, UK\label{edin}
\and Max-Planck-Institut f{\"u}r Astronomie, K{\"o}nigstuhl 17, D-69117 Heidelberg, Germany\label{mpia}
\and ETH Zurich, Institute for Astronomy, Wolfgang-Pauli-Str. 27, CH- 8093, Zurich, Switzerland\label{eth}
\and Department of Astronomy, University of Michigan, 1085 S. University, Ann Arbor, MI 48109\label{michigan}
\and Instituto de F{\'i}sica y Astronom{\'i}a, Facultad de Ciencias, Universidad de Valpara{\'i}so, Av. Gran Breta{\~n}a 1111, Valpara{\'i}so, Chile\label{valpa}
\and 
N\'ucleo Milenio de Formaci\'on Planetaria (NPF), Chile \label{nucleo}
\and DOTA, ONERA, Universit{\'e} Paris Saclay, F-91123, Palaiseau France\label{onera}
}

 \offprints{A. Boccaletti, \email{anthony.boccaletti@obspm.fr} }


\authorrunning{A. Boccaletti et al.}
\titlerunning{MWC 758}

\abstract
{Spiral arms in protoplanetary disks could be shown to be the manifestation of density waves launched by protoplanets and propagating in the gaseous component of the disk. At least two point sources have been identified in the L band in the MWC\,758 system as planetary mass object candidates.}
{We used VLT/SPHERE to search for counterparts of these candidates in the H and K bands, and to characterize the morphology of the spiral arms .}
{The data were processed with now-standard techniques in high-contrast imaging to determine the limits of detection, and to compare them to the luminosity derived from L band observations.}
{In considering the evolutionary, atmospheric, and opacity models we were not able to confirm the two former detections of point sources performed in the L band. In addition, the analysis of the spiral arms from a dynamical point of view does not support the hypothesis that these candidates comprise the origin of the spirals. }
{Deeper observations and longer timescales will be required to identify the actual source of the spiral arms in MWC\,758. }

\maketitle

\section{Introduction}
While studies of the demographics of exoplanets are well underway, the processes leading to planet formation are still poorly constrained.
Understanding how planets form is a vast topic which requires theoretical works, but the most compelling facts could be provided by the direct observation of very young systems, of a few Myrs old in age, which is precisely the moment when planets are still in the formation stage. Regardless of the difficulty in determining the ages of young systems, they are still very much embedded in their envelope. Opacity effects are critical and scale inversely with the wavelength, given that long wavelengths in the thermal regime are more appropriate for reaching the midplane of a disk, whereas short wavelengths in the scattered light regime trace the surface of flared protoplanetary disks. 

In such conditions, finding evidence of exoplanets could rely on dynamically induced structures, with observable signatures in the disk morphology. One obvious case is the presence of spiral arms, which - in the case of disks that are not especially massive - could be attributed to density waves launched by planets \citep{Goldreich1980}, but also potentially by other types of perturbations (e.g. vortices). In this context, MWC\,758 is of particular interest as a protoplanetay disk with two prominent spiral arms observed in scattered light \citep{Grady2013, Benisty2015}.  

While \citet{Benisty2015} show that reproducing the shape of the spiral arms with planets located within the 50\,au cavity would require an unphysical hot disk, \citet{Dong2015} found that outer planets with a few Jupiter masses, orbiting at ~100 au, could explain the large pitch angle. 
More recently, to match the spiral patterns in scattered light, as well as the vortices 
identified in the submillimeter \citep{Dong2018}, 
\citet{Baruteau2019} proposed two planets of 1.5 and 5\,$M_{J}$, located at 35 and 140\,au, respectively.
Finding objects responsible for the propagation of  density waves has been attempted in the L band to benefit from a lower star-planet contrast and lower dust extinction. 
Two independent studies, using Keck telescope and the Large Binocular Telescope (LBT), respectively, have reported two distinct point source candidates but with a low significance: one located at $\sim$\,20\,au from the star \citep{Reggiani2018}  that is residing inside the sub-millimeter cavity, while the second \citep{Wagner2019} is orbiting at $\sim$\,100\,au, thus showing itself to be potentially in line with the \citet{Dong2015} predictions. 
 
The purpose of this letter is to explore the detection limits provided by SPHERE \citep{Beuzit2019}, the high-contrast imaging instrument at the Very Large Telescope, based on two data sets obtained in 2016 and 2018 in the H and Ks bands. 

\section{Observations and data reduction}

\begin{table*}[th!]
\begin{center}
\begin{tabular}{l l r l r l r l r l r l r l r l r }
\hline
Date UT               &       Filter                  &         Fov rotation    &       DIT     &       N$_ \mathrm{exp}$       &       T$_\mathrm{exp}$        &     seeing                      &       $\tau_0$      & Flux var.  &       TN                      \\
                        &                                                  &       (\deg)          &       (s)     &                                       &       (s)                             &       ($''$)                          &       (ms)    & (\%)   &    (\deg)                 \\ \hline \hline
2016-01-01  &  IRDIS - K1K2 &  24.91    &   64  &   80 &    5120 &  $1.04\pm0.14$   &   $2.9\pm0.4$ & 11.2 &    $-1.750$\\
2016-01-01  &  IFS - YH     &  23.15    &   64  &   80 &    5120 &  $1.04\pm0.14$   &   $2.9\pm0.4$ &  9.6 &    $-1.750$\\ \hline
2018-12-17  &  IRDIS - H2H3 &  29.07    &   96  &   64 &    6144 &  $0.43\pm0.04$   &   $6.4\pm1.4$ & 4.3  &    $-1.764$\\
2018-12-17  &  IFS - YJ     &  29.23    &   96  &   64 &    6144 &  $0.43\pm0.04$   &   $6.4\pm1.4$ & 3.9  &    $-1.764$\\\hline
\hline
\end{tabular}
\end{center}
\caption[]{Log of SPHERE observations indicating (left to right columns): the date of observations in UT, the filters combination, the amount of field rotation in degrees, the individual exposure time (DIT) in seconds, the total number of exposures, the total exposure time in seconds, the DIMM seeing measured in arcseconds, the correlation time $\tau_0$ in milliseconds, the variation of the flux during the sequence in \%, and the true north (TN) offset in degrees. 
} 
\label{tab:log}
\end{table*}

MWC\,758 (HIP\,25793, V=8.27, H=6.56, K=5.80) was observed as part of the SPHERE survey aimed at the search for planets \citep[SHINE, SpHere INfrared survey for Exoplanets,][]{Desidera2021}. The most recent distance estimate from Gaia DR2 is $160.24\pm1.73$\,pc \citep{Gaia2018}, 
which is significantly smaller than the HIPPARCOS distance ($279^{+94}_{-58}$\pc) used in \citet{Benisty2015}, and slightly larger (but still within the error bars) than the Gaia DR1 value ($151^{+9}_{-8}$\pc) used by \citet{Reggiani2018}.

We obtained two epochs on 2016-01-01 (096.C-0241) and 2018-12-17 (1100.C-0481), using the IRDIFS-ext mode, which combines IRDIS \citep{Dohlen2008} in K1K2 (2.110, 2.251\,$\muup$m, R$\sim$20) and IFS \citep{Claudi2008} in the YH configuration (0.95-1.55\,$\muup$m, R$\sim$33), and the IRDIFS mode with IRDIS in H2H3 (1.593, 1.667\,$\muup$m, R$\sim$30); while the IFS is set in the YJ configuration (0.95-1.35\,$\muup$m, R$\sim$54). The main parameters of these observations, such as  exposure times, atmospheric conditions, and field rotation, are reported in Table \ref{tab:log}. In both observations, we used the N\_ALC\_YJH\_S coronagraph (185\,mas in diameter), which is designed for wavelengths shorter than the H band. Therefore, the IRDIS 1st epoch observation in K band suffers from a lower contrast. 

The observing sequences, and data reduction follow the standard procedure implemented in the SPHERE Data Center as described in \citet{Delorme2017}. Further details of the data reduction can be found in several papers, such as \citet{Boccaletti2018}. The conditions (seeing, coherence time) were far more appropriate for a high contrast in the second epoch in 2018. The variability of the star's  intensity (as measured during point spread function exposures at the beginning and at the end of the sequence) is an indicator of good stability ($\sim$4\% in 2018 instead of $\sim$10\% in 2016). Beside, the airmass is rather large (Am$>$1.5) for this target. 

We made use of SpeCal \citep{Galicher2018} to process the data cubes of both IRDIS and IFS with a variety of angular differential imaging (ADI) techniques. In addition, we used custom routines to perform reference differential imaging (RDI) with another star as a calibrator\footnote{Other targets observed the same night with the same observing mode, which do not necessarily match the color and magnitude of MWC\,758}, or an even simpler algorithm to perform spatial filtering.  

\section{Global description}

\begin{figure}[t!] 
\centering
\includegraphics[width=9cm]{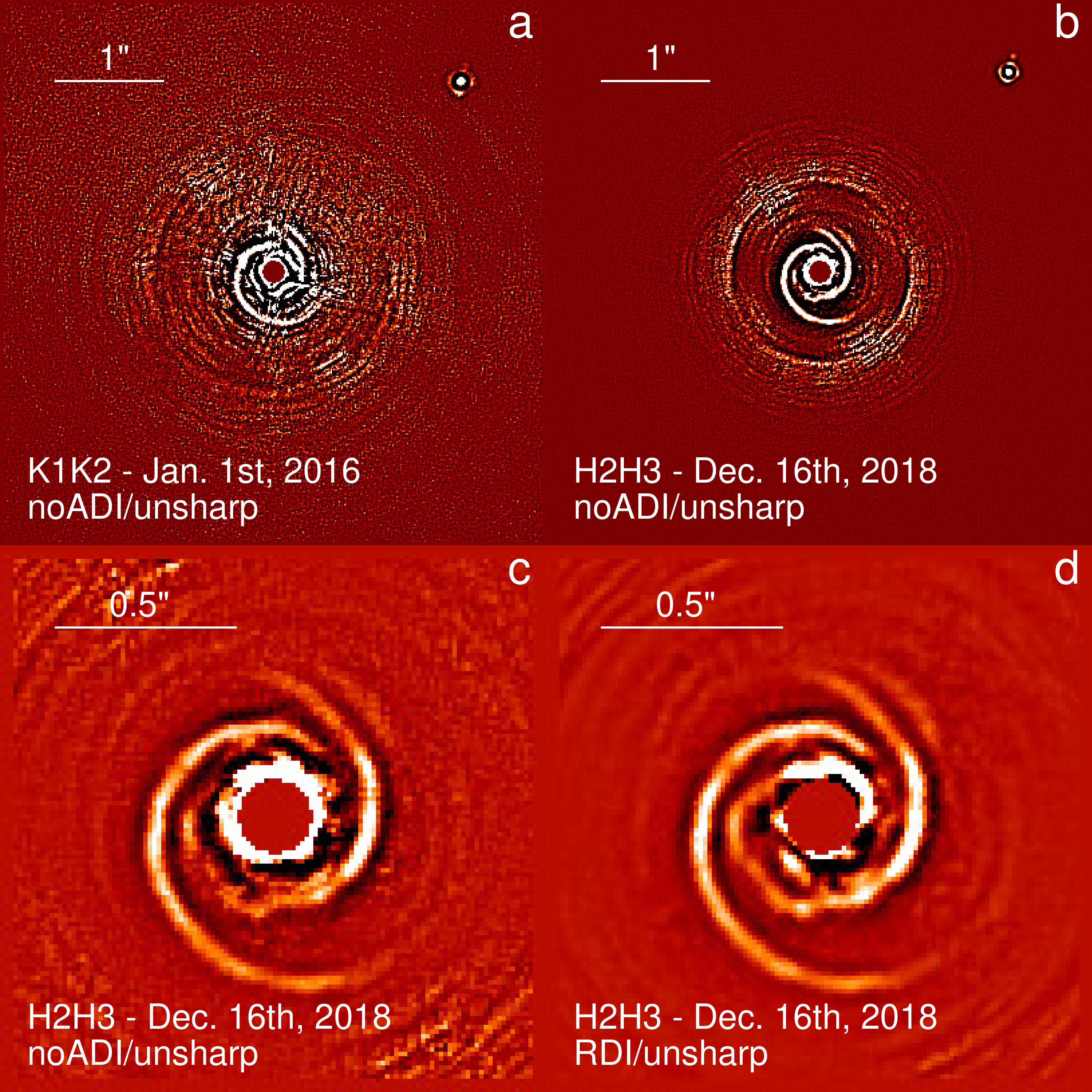}
\caption{High-pass filtering images of no-ADI and RDI data at  two epochs displayed in a large (5$''$, top) and narrower field of view (1.5$''$, bottom). North is up, east is left. The intensity scale is arbitrary.}
\label{fig:unsharp}
\end{figure}

The disk of MWC\,758 is quite bright, hence, it is already visible in raw data and stacked images, without any ADI processing  (referred as no-ADI). Subtracting the stellar residuals with the coronagraphic images of a reference star (RDI) that were obtained in similar conditions reveals the signal of the disk as close as $\sim0.1''$. Several stars were considered as references observed in the same night in the same configuration, however, here we show the one that provides the best result in terms of the detection of the disk. From these data, we measured $F_{disk}/F_\star\approx0.017$ in the H2 and H3 filters integrating the disk signal between 0.12$''$ and 0.6$''$ from the star. Figure \ref{fig:unsharp} displays the  noADI and RDI images, further processed with high-pass filtering to enhance the disk structures. The upper panels show a 5$''$ field of view, where the previously known background object \citep{Grady2005} is in fact resolved as two components, with their respective separations and position angles of  $\rho=2511.4\pm1.5$mas, PA=$316.63\pm0.03$\deg, and $\rho=2632\pm32$mas, PA=$317.9\pm0.7$\deg. The error on the second component is significantly larger because of the brightness ratio ($\sim$3.7\,mag) between the two stars and their proximity ($\sim$152\,mas), which affect the astrometric procedure. The apparent motion between the two epochs is fully consistent with the proper motion of MWC\,758.

\begin{figure*}[th!] 
\centering
\includegraphics[width=18cm]{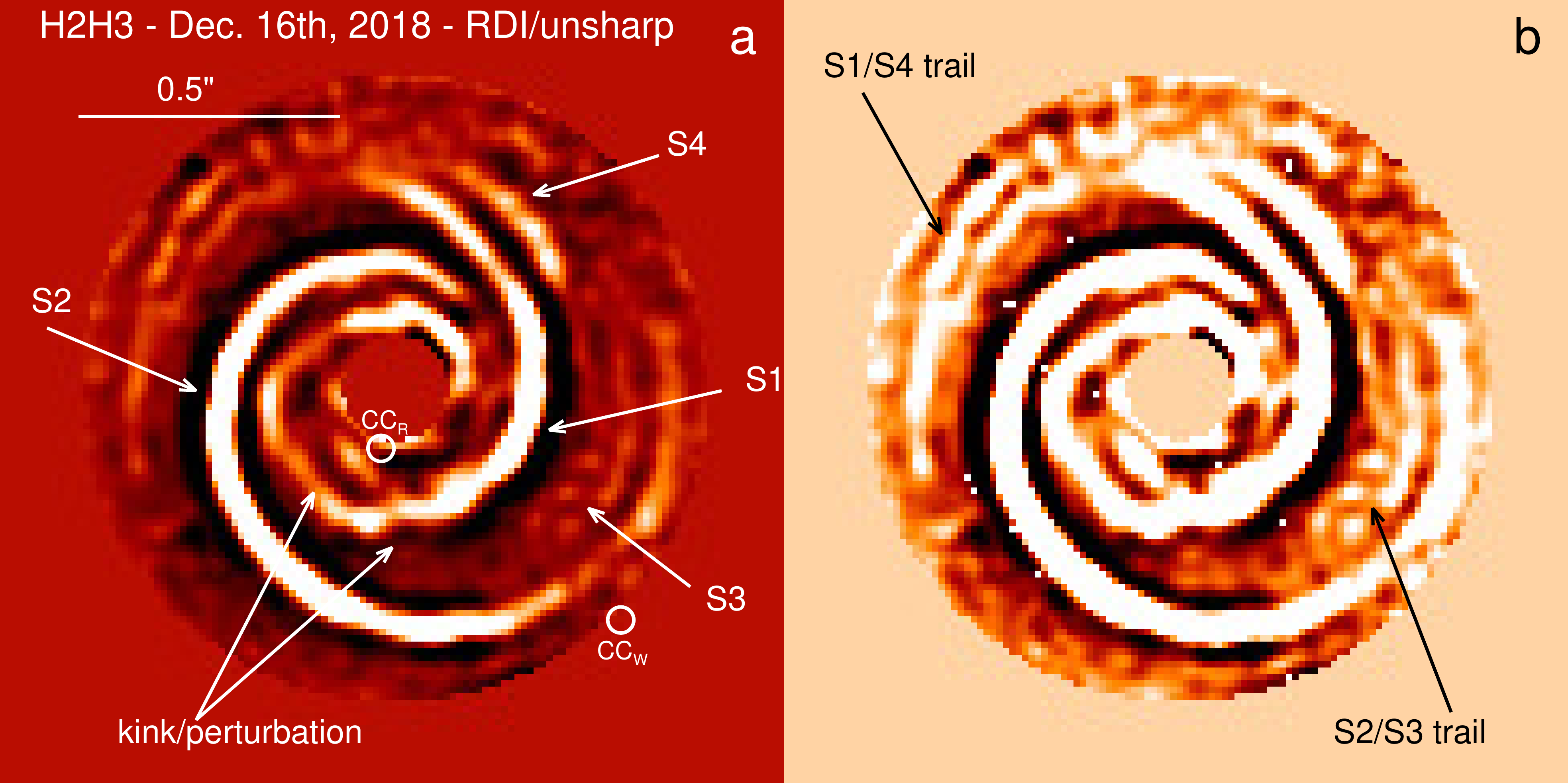}
\caption{High-pass filtering RDI images, same as Fig. \ref{fig:unsharp}d, with labels and different intensity cuts to enhance the faintest structures. North is up, east is left. The intensity scale is arbitrary.}
\label{fig:rdi}
\end{figure*}

The images contain several structures that are labeled in Fig. \ref{fig:rdi}, with the spirals S1 and S2 being the most prominent structures. It is relatively difficult to identify precisely where the spirals actually start from the shortest angular separations, but the total intensity data, especially with RDI, definitely provide a higher signal to noise ratio (S/N) of the $<0.25''$ inner region compared to the DPI image in \citet{Benisty2015} and \citet{Ren2020}.

Starting from the shortest separations,  S1 visually starts from the northeast (PA=60\deg,  $\rho=0.18''$, although it could even extend as close as the coronagraphic mask) and can be traced over one and a half rotations (Fig. \ref{fig:rdi}a). S2 originates from the northwest (PA=-30\deg, $\rho=0.22''$) and can be traced up to the west, covering a full rotation in total. While we note that it is difficult to identify exactly where the spirals start in the interior as there may be confusion with the residual starlight pattern around the coronagraphic mask, there is some evidence that the spirals extend even closer in. Thus, the RDI process (Figure \ref{fig:unsharp}d) provides a cleaner vision of the central part than the no-ADI case (Figure \ref{fig:unsharp}c).  At the other extreme, at low intensity levels, we can also identify the trails of the spirals  (Fig. \ref{fig:rdi}b). This is by far the deepest observation of these spiral arms. 
We also recovered some structures in the southwest of S1 (at $PA = 235-290\degb$), labeled S3, which could be reminiscent of a feature identified by \citet{Reggiani2018}. 
The SPHERE data presented here convey the possibility that S3 and S4 could be connected (S4 being in the trail of S3). This possible link could have been already posited from the L band data in \citet{Reggiani2018} and has already been established by \citet{Calcino2020} based on hydrodynamical simulations.
Thus, S4 could also be considered as the bottom side of the disk in which case the angular separation between S1 and S4, given the inclination of 21$\degb$ is related to the disk opening angle which we estimated to $\sim$23$\degb$, indicative of a rather large disk thickness.

While S2 is apparently close to an Archimedean spiral, it is not the case for S1, which appears perturbed in the  PA range 60-180\deg, with a sort of kink near the location: PA$\approx 176$\deg, $\rho\approx 0.25''$,  possibly splitting into two components (Fig. \ref{fig:rdi}b). Although the RDI processing has lower impact on the disk morphology than ADI, the exact distribution of the structures at such stellocentric distances remain to be confirmed. We also present the IFS RDI processed images (Fig. \ref{fig:ifs}) which provide a much lower signal to noise ratio than with IRDIS.

When processed with ADI (Fig. \ref{fig:adi}), the spirals are no longer visible because of the small field rotation ($<$30\deg) which induces a quite strong self-subtraction \citep{Milli2012}. Instead, many structures (consistent with the kink and perturbation seen in RDI) are revealed along the spirals, some of which could be confused with the usual signatures of point sources in ADI. These patterns are consistent across wavelengths between IRDIS and IFS (Fig. \ref{fig:adi}, b and d for instance). 
Due to the difference in starlight rejection between the two epochs obtained $\sim$3 years apart, it is still difficult to perform a thorough analysis of the evolution of these structures, especially within the sub-millimeter cavity ($<0.3''$).

\section{Limits of detection to point sources}

The limits of detection are estimated with \texttt{SpeCal}, as explained in \citet{Galicher2018}. Here, we made use of the KLIP algorithm \citep{Soummer2012} which produces a higher contrast when compared to other types of ADI implementations in this particular case. The contrast at a given radius is calculated from the standard deviation of the signal contained in an annulus of 0.5$\times$FWHM in width (about 2 pixels) centered on the star. 
The self-subtraction induced by ADI is estimated with fake planets injected into the datacube (along a spiral pattern to cover a range of separations and azimuths). The presence of the disk's spirals and their residuals after ADI processing inevitably corrupts this measurement. 
Both the self-subtraction and the radial transmission of the coronagraph \citep[as reported in ][]{Boccaletti2018} are taken into account  to produce the final contrast plots in Fig \ref{fig:limdet} (left). The second epoch is significantly better with, for IRDIS, a gain of about one order of magnitude in contrast, in the separation range of $0.1''-0.2''$ (as a result of a sub-optimal coronagraph used in K band and poorer conditions), and a more modest gain of 2 to 4 further out. The gain is less pronounced but still noticeable for the IFS sequences. 

\begin{figure}[t!] 
\centering
\includegraphics[width=9cm]{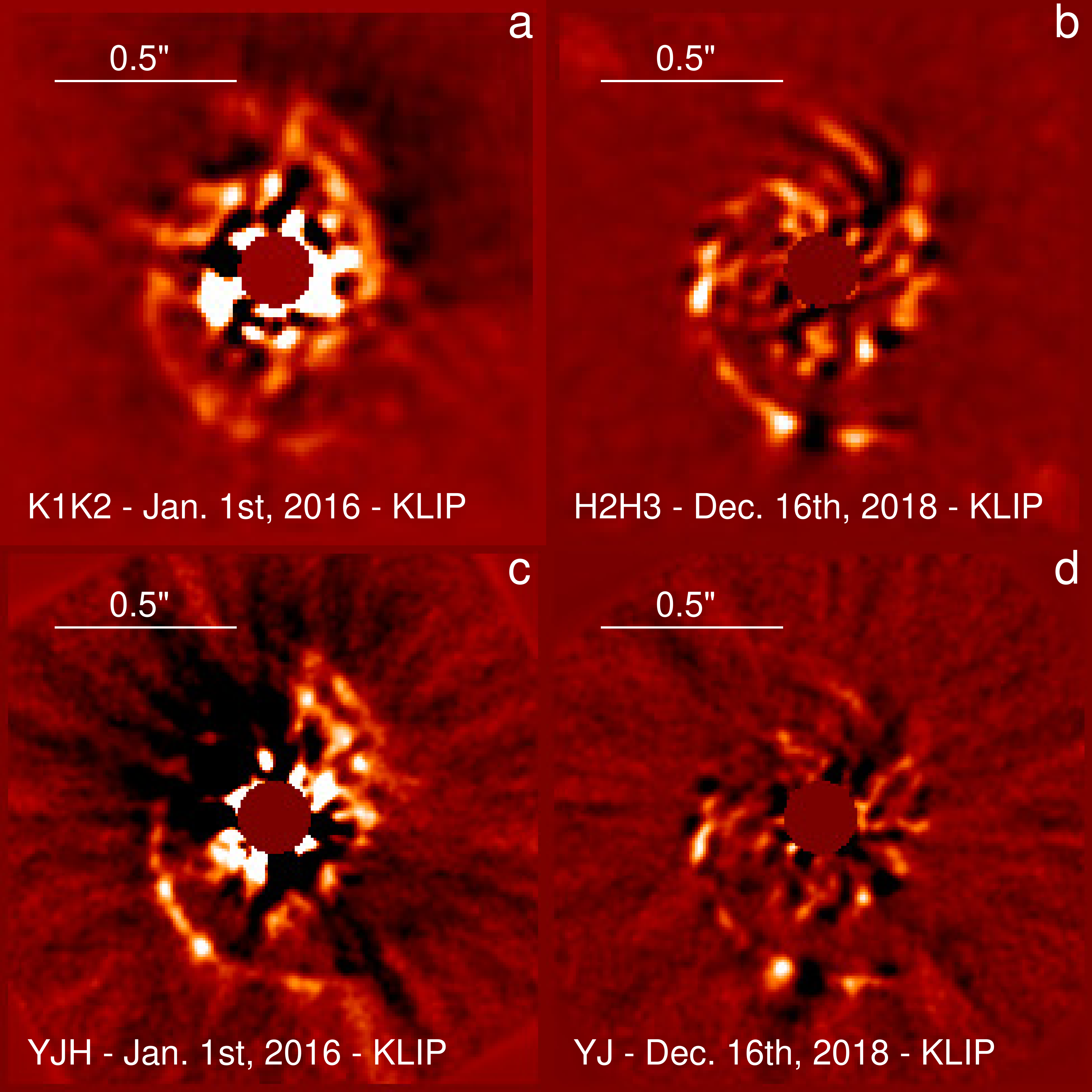}
\caption{ADI processing of the two epochs in a 1.5$''$ field of view for IRDIS (top) and IFS (bottom). North is up, east is left. The intensity scale is arbitrary.}
\label{fig:adi}
\end{figure}

Two point sources were identified from previous L band observations. The first companion candidate (CC$_R$) reported by \citet{Reggiani2018} using Keck is located very close to the star inside the cavity at $\rho=0.112\pm0.006''$, $PA=169\pm4\degb$, and with a brightness ratio of $\Delta L'=7.1\pm0.3$. The second point source candidate (CC$_W$) was observed with the LBT by \citet{Wagner2019} along the southern spiral at $\rho= 0.617\pm0.024''$, $PA= 224.9\pm2.2\degb$, and $\Delta L'=12.5\pm0.5$. 
The locations of these candidate point sources are reported in Fig. \ref{fig:rdi}a and Fig. \ref{fig:spirals}b (not taking into account orbital motions which are presumably small). 
It is important to note that none of these two observations confirms the other, although this can be a matter related to the signal-to-noise ratio. Indeed, the spiral patterns of the disk are visually better detected in \citet{Reggiani2018} images.
\citet{Wagner2019} claimed CC$_W$ to be of planetary nature, and its measured $L'$ band photometry would correspond to a mass of 2 to 5\,$M_{J}$ assuming an age of $1.5-5.5$\,Myr. Using the COND atmosphere models \citep{Allard2001}, as in \citet{Wagner2019}, along with an age estimate of 1 to 5 Myr identical to the one used by \citet{Reggiani2018, Wagner2019} , the contrast performance of SPHERE at 0.6$''$ measured in the H band translates into a mass sensitivity of $0.6 - 1.5\,M_J$ (Fig. \ref{fig:limdet}, right). The K band data are less constraining as the achieved contrast corresponds to a mass range of $1.9 - 4.5\,M_J$. Therefore, the SPHERE observations do not support the presence of a point source at the location of CC$_W$, nor the detection limits are compatible with the expected mass range. 
Moreover, it is obvious in Fig. \ref{fig:spirals}b that CC$_W$ is not located along the trace of the spiral arm with a radial departure of about $0.08''$ that is twice the angular resolution of the SPHERE images. Given the level of clumping that is observed from the innermost separations along the spirals and all the way out (Fig. \ref{fig:adi}), CC$_W$ could be a dust feature (although it should be also visible at shorter wavelengths in reflected light), or an artifact given the low signal-to-noise ratio reported in \citet{Wagner2019}.

On the contrary, for dynamical reasons, \citet{Reggiani2018} argue that the emission from the closest point source, CC$_R$, cannot originate from its photosphere or it would translate to a mass of 41 to 64\,$M_J$; hence, in the brown dwarf regime and would have induced a noticeable cavity in the small dust grains distribution. Such a massive companion would have been presumably detected with SPHERE, according to Fig. \ref{fig:limdet} (right). Instead, the authors postulate that the emission comes for a circumplanetary disk (CPD), 
the luminosity of which depends on the product of the planet mass ($M_p$) and the accretion rate ($\dot{M}$). Assuming face-on 1-Myr old disks and constant accretion rate, \citet{Zhu2015} tabulated the absolute magnitudes in near-IR filters for several values of the accretion disk inner radius $R_{in}$ (from 1 to 4 Jupiter radii). Under this theoretical framework, the photometry of CC$_R$ in the L band would correspond to an object of $\sim1-8\,M_J$ for a moderate accretion rate of $10^{-8}M_{\odot}$/yr \citep[consistent with the limit of detection in H$\alpha$ reported by][]{Huelamo2018}. 
As a comparison, from the L band photometry we can derive the expected absolute magnitude in the H band for such a CPD according to  \citet[][Table 1]{Zhu2015}. We found a value ranging from H=7.7 to 10.3 ($R_{in}=1-4$), while we measured a contrast of $\Delta H=9.57$, hence, obtaining an absolute magnitude of $H_{limdet}= 10.11$. Therefore, the non-detection of a point source in SPHERE data at the position of CC$_R$ is only marginally consistent with this flux having been produced by a CPD. In fact, this assumption is even less consistent with a non-detection in the K band since $K_{limdet}= 8.78$ ($\Delta K=9.00$) and the L band photometry of a CPD should correspond to $K = 7.08 - 8.35$.

\section{Extinction by the disk}
Implicitly, the assessment of detection limits in the previous section assumes the same optical depth at all wavelengths. In a more realistic case, the surrounding disk is expected to produce some extinction and it is necessary to estimate the variation of optical depth between the L band, and the H and Ks bands to derive meaningful conclusions. The optical depth can be assessed using an estimate of the surface density at the position of the companion candidates and a model of the dust opacity.
\citet{Boehler2018} considered the disk surface brightness of MWC\,758 measured in the continuum with ALMA, and coupled to a radiative transfer model, to derive azimuthally averaged surface density values of the order of $4.4\,10^{-3}$ ($3.3\,10^{-3}$ g/cm$^2$ resp.) at the position of CC$_W$ (CC$_R$ resp.). 
Using a dust model in which the composition and the size distribution are consistent with parameters used by \citet{Boehler2018}\footnote{The mass fractions are 0.257/0.18/0.563 for silicates, amorphous carbon, and water ice respectively; the size distribution starts from $a_{min}=0.05\,\muup$m to $a_{max}=1$\,mm with a -3.5 power-law exponent, and the dust porosity is 20\% \citep{Woitke2019}}, 
we computed dust mass-opacities in the filters of interest, in the H, K, and L' bands. Combined with the half surface densities estimates (to consider only the amount of dust mass from the disk midplane to the surface) at CC$_R$ and CC$_W$ locations, we calculated the difference of magnitudes, which are displayed in Tab. \ref{tab:deltamags}. 
However, since water ice is absent from many dust models of protoplanetary disks \citep[][and this is particularly relevant for the surface layers of disks where most of the opacity in the H/K/L' bands is produced]{Woitke2019}, we also computed these magnitude differences when the water ice component  is removed. The differences of magnitudes  (H-L' and K-L') can even become negative due to some extinction resonances at specific wavelengths. In the case of the "standard" particle size distribution starting at $a_{min}=0.05 \,\muup$m, the magnitude differences between the H/K and L' bands are small. In the case of a larger minimum size ($a_{min}=1 \,\muup$m), which would correspond to a more evolved/processed dust population, we observe more dispersion in the magnitudes differences. 
Yet, all the magnitudes extinctions remain below one magnitude, which shows that a large amount of differential extinction is unlikely - an argument with the potential to reconcile present SPHERE observations and previous ones. Furthermore, CC$_R$ and CC$_W$ inferred companions are massive enough to carve gaps in the disk, which would  reduce the dust opacity effects even more. 

\begin{table}
    \begin{center}
    \begin{tabular}{l c c c c}
    \hline \hline
    Dust min. size & \multicolumn{2}{c}{CC$_R$}	& \multicolumn{2}{c}{CC$_W$} \\
                   &     H-L'   	            &	      K-L'   &    H-L'   &   K-L'	 \\
	($\muup$m)       &     (mag)       &     (mag)    &   (mag) 	 &      (mag)      \\
	\hline \hline
	 \multicolumn{4}{c}{silicates+carbon+water ice dust} \\
     0.05       &  0.02	 &  -0.16 	& 0.03 	 & -0.22  \\
     1.0        &  -0.42	 &  0.39 	& -0.57 	 & 0.51  \\
     \multicolumn{4}{c}{silicates+carbon dust} \\
     0.05       &  0.62	 &  0.26 	& 0.83 	 & 0.34  \\
     1.0        &  -0.10	 &  0.23 	& -0.13 	 & 0.30  \\
    \end{tabular}
        \caption{Magnitude differences with respect to the L' band estimated at the positions of $CC_R$ and $CC_W$ for two assumptions on the minimum grain size. }
    \label{tab:deltamags}

\end{center}
\end{table}

The detection limits derived in the previous section can be re-visited given the extinctions estimated in Table \ref{tab:deltamags} and in considering the worst-case scenario (one of the two dust compositions). Assuming the H band detection limit would suffer from differential extinction of 0.83 magnitude (resp., 0.34 magnitude in K band) at the location of CC$_W$ for the smallest minimum grain size, the mass limit increases to $0.9 - 2.0\,M_J$ (resp., $2.2-5.3\,M_J$ in K band), which is still lower than (resp., comparable to) the mass inferred by \citet[][$2 - 5\,M_J$]{Wagner2019}. Larger minimum grain sizes imply either a lower extinctions in H (-0.13 magnitude) or slightly larger in K (0.51 magnitude), which does not, therefore, change drastically the outcome. 

In what concerns the object CC$_R$, the extinction moves the limit of sensitivity from $H_{limdet}=10.11$, to 9.49 for $a_{min}=0.05 \,\muup$m, respectively 10.53 for $a_{min}=1 \,\muup$m. In the K band, the sensitivity reaches 8.52 or 8.39 instead of 8.78 depending on the minimum grain size.
Comparing with the extrapolated magnitudes of a CPD in the H band ($H=7.7 - 10.3$) and the K band ($K=7.08-8.35$), these limits of sensitivity are sufficient to rule out - either partially (H band) or totally (K band) - the presence of a CPD.

We note that at the position of CC$_W$ and CC$_R$, the local surface density in sub-micronic/micronic particles could be different from that extrapolated from \cite{Boehler2018}, with ALMA measurements tracing the larger ones. Also, additional accretion along planet poles could also make the extinction higher than evaluated, especially in a face-on geometry \citep{Fung2016}.

\section{Spirals as tracers of point sources}
We measured the spine of the two main spirals (S1 and S2) following the method developed in \citet{Boccaletti2013}, along with an adequate numerical mask to isolate each spiral (data points are reported in Fig. \ref{fig:spirals}, left). 
The spirals are relatively well fitted with an archimedean function ($\rho=a\times\theta+b$) in the PA range $[-70\degb,330\degb]$ for S1 (green circles, $\chi_{\nu}^2=1.27$, $a=0.29\,mas/\degb$, $b=159\,mas$) and  $[-30\degb, 210\degb]$ for S2 (blue circles, $\chi_{\nu}^2=1.20$, $a=1.11\,mas/\degb$, $b=128\,mas$).

Planetary mass objects in the MWC\,758 system can be responsible for the launching of density waves materializing into spiral arms in and out of the planet orbital radius. 
An obvious test would be to check whether we can associate the spirals S1 and S2 with CC$_R$ and CC$_W$. However, a thorough investigation will require dedicated hydrodynamical simulations in 3D which is beyond the scope of this paper. Previous works extensively used the linear approximation to the density waves theory \citep{Rafikov2002} to interpret the geometry of spiral arms observed in scattered light \citep{Muto2012, Boccaletti2013, Benisty2015} although this approach has been found unreliable for massive planets and for cases when the spiral wake is localized away from the disk plane, at the disk surface, as in the case for scattered light images \citep{Zhuetal2015}.  
Nevertheless, the pitch angle of the outer spiral arm is reduced by the disk 3D structure which compensates for the wake broadening. Hence, the linear theory still provides a decent match for the outer spiral even for a massive planet (several Jupiter masses).  
With these limitations in mind we performed a qualitative analysis following the prescription of \citet{Muto2012}: 
\begin{eqnarray}
 \theta(r) &=& \theta_0 + \frac{\text{sgn}(r-r_c)}{h_c}       \nonumber \\
   	&&\times  \left[  \left(\frac{r}{r_c}\right)^{1+\beta} \left\{ \frac{1}{1+\beta}-\frac{1}{1-\alpha+\beta} \left(\frac{r}{r_c}\right)^{-\alpha}  \right\} \right. \nonumber\\
   && \left. -\left( \frac{1}{1+\beta} - \frac{1}{1-\alpha+\beta} \right)\right]
   \label{eq:muto}
\end{eqnarray}

where $h_c$ is the disk aspect ratio at the planet location ($r_c$, $\theta_0$), $\alpha$ and $\beta$ are the power-law exponents of the angular frequency and temperature profile dependence with $r$ (in the standard case $\alpha=1.5$, $\beta=0.25$). In addition, we assumed the disk inclination and position angle: $i=21\degb$ and $PA=62\degb$, together with $h_c=0.18$ \citep{Boehler2018, Andrews2011}. We also accounted for a standard flaring index $\gamma=1.2$.  
Imposing the locations of the two candidate companions CC$_R$ and CC$_W$, we calculated the associated spirals launched from these sites (respectively the green and blue lines in Fig. \ref{fig:spirals}, right). 

Focusing on the outer arms of the spirals, beyond the planets orbital radii, 
we found no obvious match between these spiral models and those observed in the data. 
This may suggest that the perturber(s) could be much less massive or they could be located at smaller physical distances (or both). 
Again, we caution that the density wave linear theory does not capture the complexity of this system and cannot be taken as a definitive result in this case. 
A more elaborate modeling is proposed in \citet{Calcino2020}, which qualitatively reproduces the main disk features assuming a 10\,M$_J$ planet at 33.5\,au (about $0.2''$) with some eccentricity ($e=0.4$).

As long as planet-induced spirals are co-rotating with the planets, we can also consider dynamical arguments to evaluate the amount of expected rotation of the spirals. For an object at the position of $CC_R$, and $CC_W$, we expect an angular motion between the two epochs (2.96 years apart) of $\sim17.9\degb$,  $\sim1.3\degb$ respectively, for a central star with a mass of 1.7M$_\odot$ (average value in the literature). Therefore, we can safely reject $CC_R$ as being the source of the spiral arms, as also derived by  \citet{Ren2020}. To obtain a more precise estimate of the spiral rotation, we isolated the spiral S2 with a numerical mask in the high-pass filtered RDI images from Fig. \ref{fig:unsharp} (we ignore S1 as being significantly contaminated by diffraction residuals in Jan. 2016 data), deprojected the images, multiplied by the square of the stellocentric distance to give more weight to the outer parts of the spiral, and then we solved for the rotation angle which minimizes the difference between the two epochs with a least square metric. While the exact value depends on the image processing (masking and radial weighting, in particular), we can reasonably constrain the rotation to $<0.3\degb$ on the 2.96 years time frame, which means we can exclude a planet at a distance shorter than $\sim$280\,au for being a candidate for the launching of the spirals. Such an upper limit is about four times smaller than the expected rotation induced by $CC_W$. We note that this is slightly lower than the value derived by \citet[][$0.22\degb\pm0.03\degb$/yr]{Ren2020}, who carefully estimated the error budget (but it is still not compatible with $CC_W$).

\section{Conclusion}

We  imaged the system MWC\,758 with SPHERE as part of the near-IR survey at two epochs in Jan. 2016 and Dec. 2018. The data were processed with angular differential imaging to optimize the sensitivity to point sources, as well as with reference differential imaging to determine the morphology of the spirals. We did not recover the previously detected point sources reported by \citet{Reggiani2018} and \citet{Wagner2019}, nor did we detect other potential companions. Since  former detections were obtained at longer wavelengths, namely, in the L band, in which the self-luminous objects of interest (planet photospheres or circumplanetary disks) are expected to be brighter than in SPHERE band passes, we extrapolated their magnitudes to the H and K bands and compared them against the SPHERE limits of detection. We also accounted for the extinction arising from the presence of dust based on ALMA observations. Overall, the SPHERE data may rule out the presence of these candidate companions, especially in the K band (and although contrast performance is worse than in the H band). 
In addition, no satisfactory solution can be found with linear density-wave models of the spirals arms that assume the spiral-driving planets are located at the position of CC$_R$ and CC$_W$.
Finally, the amount of rotation of the spirals between the two epochs also sets stringent limit on the stellocentric distance of planetary mass companions shaping the spirals. Therefore, at this stage, pending methods than would allow greater contrasts can be reached - or finer dynamical studies, the SPHERE observations based on the near-IR survey fails to identify point sources as the source of the MWC\,758 spiral arms.

\bibliographystyle{aa}
\bibliography{mwc758}

\begin{appendix}
\section{IFS RDI images}
\begin{figure*}[t!] 
\centering
\includegraphics[width=18cm]{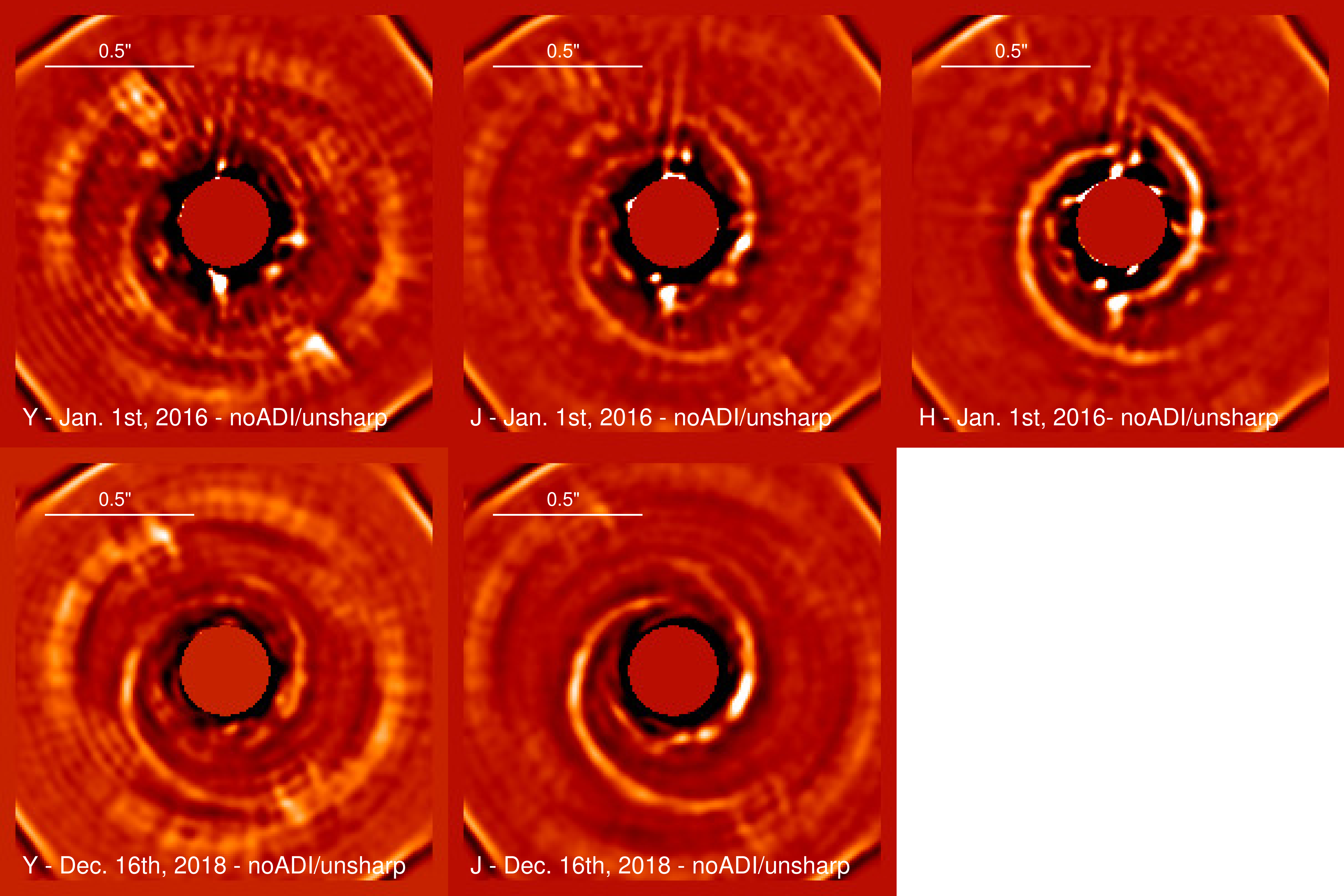}
\caption{RDI-processed images obtained with IFS data, collapsed in three channels, YJH for Jan. 2016 (top), and two channels YJ for Dec. 2018 (bottom).}
\label{fig:ifs}
\end{figure*}

\section{Limits of detection}
\begin{figure*}[t!] 
\centering
\includegraphics[width=9cm]{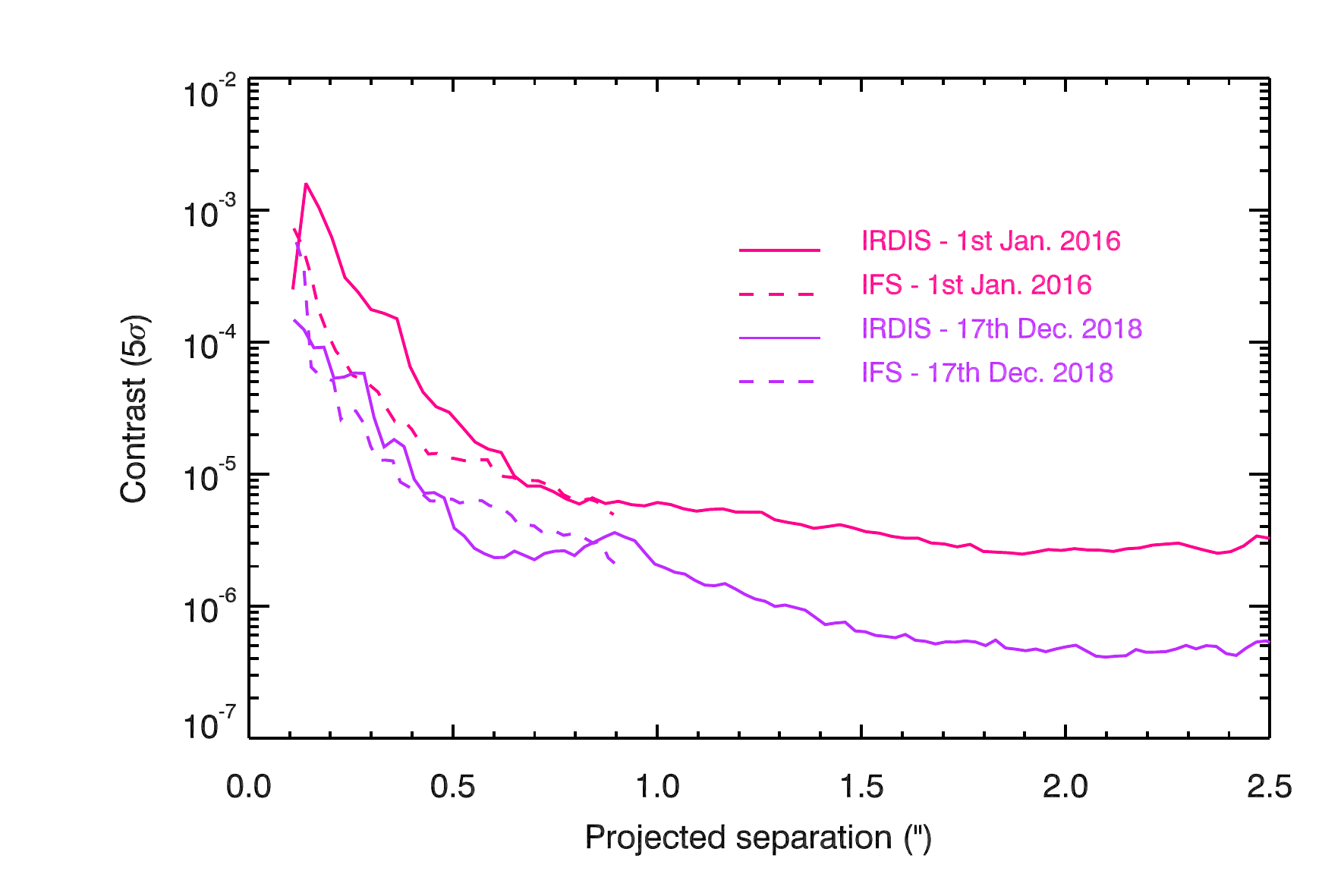}
\includegraphics[width=9cm]{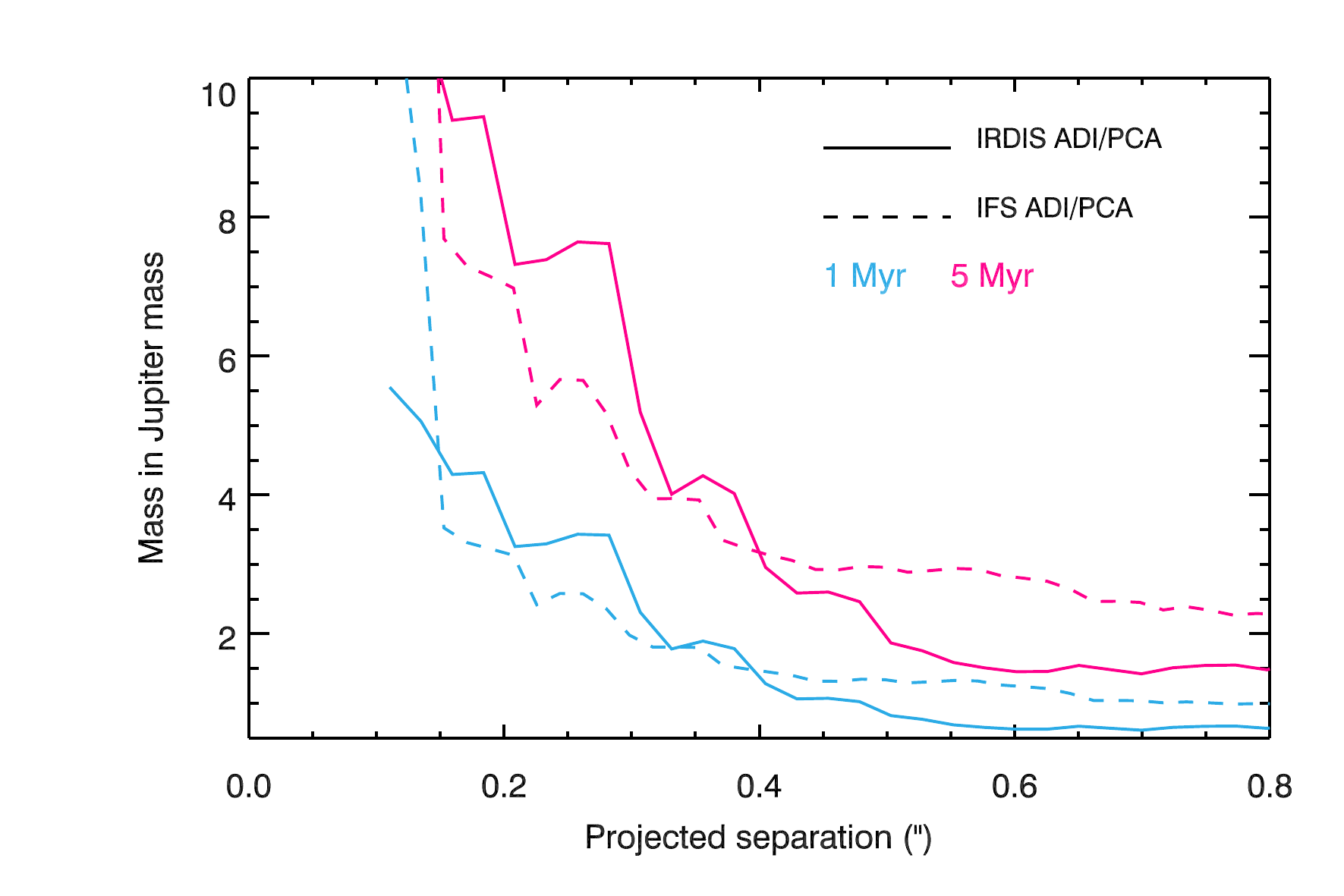}
\caption{Limits of detection in contrast (left) for both IRDIS (solid line) and IFS (dashed line) and for the two epochs (Jan. 2016, Dec. 2018). The conversion to Jovian masses (right) starts from the best contrast achieved in Dec. 2018, assumes two ages of respectively 1 and 5 Myr and uses the COND atmosphere model.}
\label{fig:limdet}
\end{figure*}

\section{Models of spiral arms}
\begin{figure*}[t!] 
\centering
\includegraphics[width=18cm]{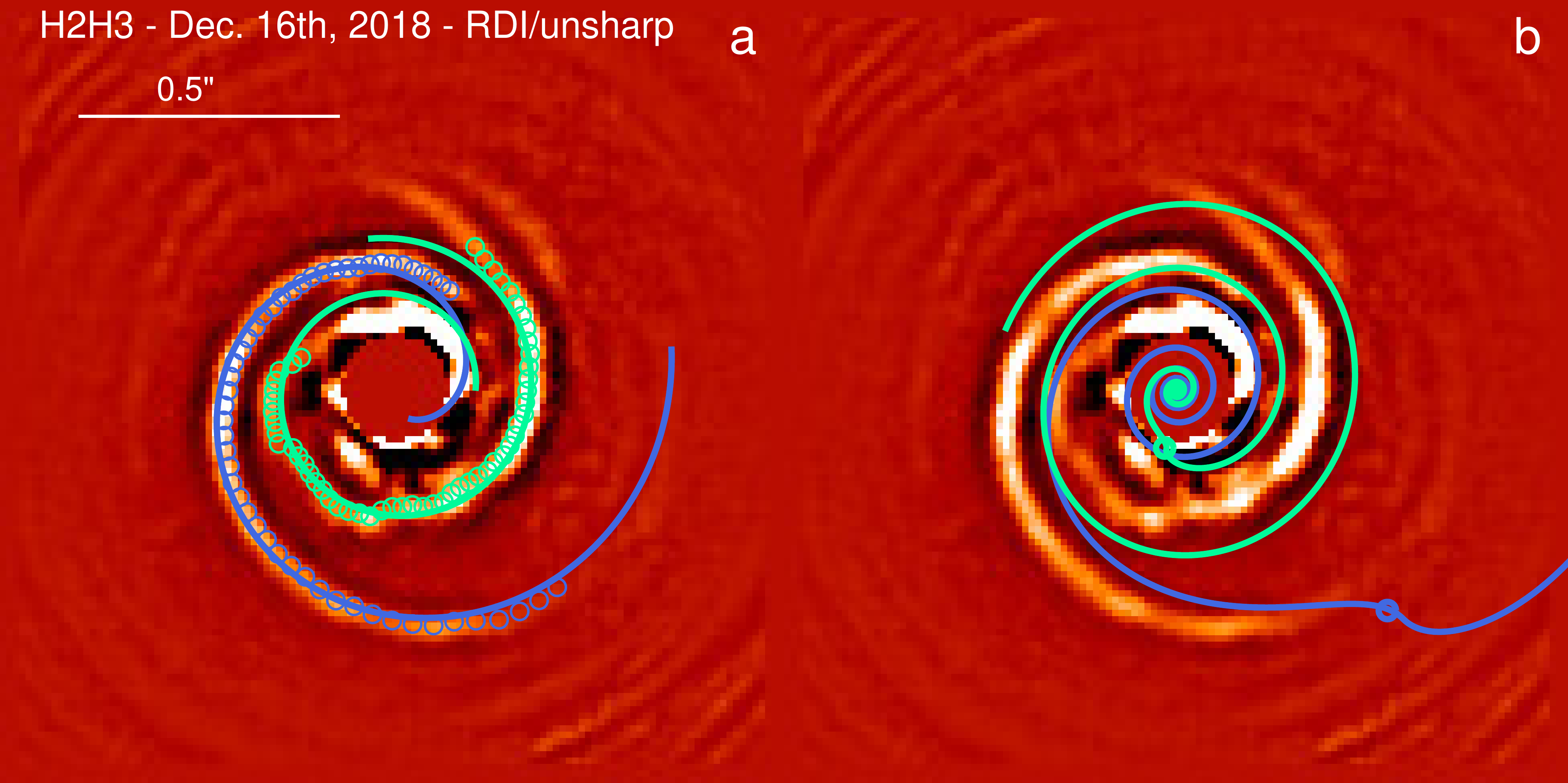}
\caption{Spines of the spirals S1 (green circles) and S2 (blue circles) superimposed on RDI processed image from Dec. 2018 together with the fit of archimedean spirals (left panel). The right panel shows two spiral models with $CC_R$ (green line) and $CC_W$ (blue line) as the perturbers, assuming linear density wave theory.}
\label{fig:spirals}
\end{figure*}
\end{appendix}

\begin{acknowledgement}
SPHERE is an instrument designed and built by a consortium consisting of IPAG (Grenoble, France), MPIA (Heidelberg, Germany), LAM (Marseille, France), LESIA (Paris, France), Laboratoire Lagrange
(Nice, France), INAF–Osservatorio di Padova (Italy), Observatoire de Genève (Switzerland), ETH Zurich (Switzerland), NOVA (Netherlands), ONERA (France) and ASTRON (Netherlands) in collaboration with ESO. SPHERE was funded by ESO, with additional contributions from CNRS (France), MPIA (Germany), INAF (Italy), FINES (Switzerland) and NOVA (Netherlands).  SPHERE also received funding from the European Commission Sixth and Seventh Framework Programmes as part of the Optical Infrared Coordination Network for Astronomy (OPTICON) under grant number RII3-Ct-2004-001566 for FP6 (2004–2008), grant number 226604 for FP7 (2009–2012) and grant number 312430 for FP7 (2013–2016). 
French co-authors also acknowledge financial support from the Programme National de Planétologie (PNP), the Programme National de Physique Stellaire (PNPS), and the programme national "Physique et Chimie du Milieu Interstellaire(PCMI) of CNRS-INSU in France with INC/INP co-funded by CEA and CNES. 
This work has also been supported by a grant from the French Labex OSUG@2020 (Investissements d’avenir – ANR10 LABX56). The project is supported by CNRS, by the Agence Nationale de la Recherche (ANR-14-CE33-0018).
Italian co-authors acknowledge support from the "Progetti Premiali" funding scheme of the
Italian  Ministry  of  Education,  University,  and  Research.
This work has been supported by the project PRIN-INAF 2016 The Cradle of Life - GENESIS-
SKA (General Conditions in Early Planetary Systems for the rise of life with SKA). 
C. P. acknowledge financial support from Fondecyt (grant 3190691) and support by ANID, -- Millennium Science Initiative Program -- NCN19\_171
Finally, this work has made use of the the SPHERE Data Centre, jointly operated by OSUG/IPAG (Grenoble), PYTHEAS/LAM/CESAM (Marseille), OCA/Lagrange (Nice) and Observatoire de Paris/LESIA (Paris). We thank P. Delorme and E. Lagadec (SPHERE Data Centre) for their efficient help during the data reduction process. 

\end{acknowledgement}

\end{document}